\documentclass[aip,graphicx,reprint]{revtex4-1}
\usepackage[cmex10]{amsmath}
\usepackage{ amssymb }
\usepackage{graphicx}
\usepackage{float}
\usepackage{multirow}
\usepackage{epstopdf}
\usepackage{color}
\usepackage{textcomp}
\usepackage[margin=0.6in]{geometry}

\draft 

\begin{document}

\title{Nano strain-amplifier: making ultra-sensitive piezoresistance in nanowires possible without the need of quantum and surface charge effects} 

\author{Hoang-Phuong Phan}
\email{hoangphuong.phan@griffithuni.edu.au}
\affiliation{Queensland Micro- and Nanotechnology Centre, Griffith University, Queensland, 4111, Australia}
\author{Toan Dinh}
\affiliation{Queensland Micro- and Nanotechnology Centre, Griffith University, Queensland, 4111, Australia}
\author{Takahiro Kozeki}
\affiliation{Department of Mechanical Engineering, University of Hyogo, 671-2280, Japan}
\author{Tuan-Khoa Nguyen}
\affiliation{Queensland Micro- and Nanotechnology Centre, Griffith University, Queensland, 4111, Australia}
\author{Afzaal Qamar}
\affiliation{Queensland Micro- and Nanotechnology Centre, Griffith University, Queensland, 4111, Australia}
\author{Takahiro Namazu}
\affiliation{Aichi Institute of Technology, Aichi, Toyota, 470-0392, Japan}
\author{Nam-Trung Nguyen}
\affiliation{Queensland Micro- and Nanotechnology Centre, Griffith University, Queensland, 4111, Australia}
\author{Dzung Viet Dao}
\affiliation{Queensland Micro- and Nanotechnology Centre, Griffith University, Queensland, 4111, Australia}


\begin{abstract}
This paper presents an innovative nano strain-amplifier employed to significantly enhance the sensitivity of piezoresistive strain sensors. Inspired from the dogbone structure, the nano strain-amplifier consists of a nano thin frame released from the substrate, where nanowires were formed at the centre of the frame. Analytical and numerical results indicated that a nano strain-amplifier significantly increases the strain induced into a free standing nanowire, resulting in a large change in their electrical conductance. The proposed structure was demonstrated in p-type cubic silicon carbide nanowires fabricated using a top down process. The experimental data showed that the nano strain-amplifier can enhance the sensitivity of SiC strain sensors at least 5.4 times larger than that of the conventional structures. This result indicates the potential of the proposed strain-amplifier for ultra-sensitive mechanical sensing applications.   
\end{abstract}


\maketitle 

Strain sensors have been widely employed in numerous applications including bio-analysis, inertia sensing, and structural health monitoring (SHM) \cite{Barlian,phan1,phan2,pressure}. For instance, in SHM, strain sensors can detect crack generation, delamination between layers, and thermal expansions due to the changes in temperature \cite{SHM1,SHM2}. Among several methods to detect strain, the piezoresistive effect in semiconductors has been widely adopted due to its high sensitivity and simple readout circuitry \cite{Piezo1,Piezo2,Piezo3,Piezo4}. 

Recent studies have been focusing on enhancing the sensitivity of piezoresistive strain sensors by down-scaling piezoresistive elements to a nanometer scale \cite{Rowe1}. 
He and Yang reported a giant longitudinal piezoresistive coefficient of $-3550\times 10^{-11}$ Pa$^{-1}$ in silicon nanowires, which is at least one order of magnitude large than that of bulk Si material \cite{Peidongyang}. The enhancement of the piezoresistive effect in Si nanowires was hypothesized to be caused by a piezopinch phenomenon \cite{Rowe2}. Following the work of He and Yang, a large number of studies have been carried out to investigate the piezoresistive effect of nanowires fabricated using different methods and aligned in several crystallographic orientations.  Milner \emph{et al.} reported the giant piezoresistive effect in Si micro and nano wires fabricated using a top-down process \cite{Milne}. In addition, the authors also counteracted the hypothesis of the piezopinch phenomenon, and suggested that the dynamic properties of surface charge on micro/nanowires could be the main reason causing the significant change in the piezoresistance of nanowires \cite{Rowe1,Milne}. Nevertheless, the electrical conductance of Si nanowires, using the dynamic properties of surface state, varies with time, which is not a desirable property for practical strain sensing applications. Additionally, in contrast to the results of He and Yang, the piezoresistive effect in both bottom-up grown Si nanowires \cite{Lugstein,CJHuang}, and top-down fabricated Si \cite{Toriyama1,Toriyama2,Phan_SiNW} reported recently, did not show significant improvement in sensitivity compared to when bulk materials are used. In another study, Nakamura \emph{et al.} theoretically investigated the influence of the quantum confinement on the piezoresistive effect in ultra narrow Si nanowires \cite{Nakamura}. Although, theoretical calculation showed a giant piezoresistive effect in nanowires, to make the quantum confinement effective, the diameter of nanowires has to be below a few nanometers, which is relatively challenging to fabricate. Therefore, to date, the existence of the large piezoresistive effect in nanowires using electrical approaches is still a controversial topic, and further studies need to be carried out to verify these methods \cite{Rowe1}.


In this paper, we report a revolutionary mechanical approach to enhance the sensitivity of piezoresistive strain sensors using a nanowire-based strain amplifying structure (hereafter nano strain-amplifier). The nano strain-amplifier was inspired from the dogbone structure, in which strain can be magnified in the desired areas. For the purpose of demonstration, we developed and characterized a p-type cubic silicon carbide (3C-SiC) based nano strain-amplifier. Silicon carbide was selected in this work due to its high potential in applications for harsh environments \cite{Senesky,Toan}, including strain sensing devices for structural health monitoring \cite{SiCNW1,SiCNW2,SiCNW3,SiC_FIB,SiCNW1_Phan}. The experimental data shows that, by using the nano strain-amplifier, the sensitivity of SiC based strain sensor can increase approximately 6 fold than that of conventional SiC micro and nano structures. Furthermore, by employing the proposed nano strain-amplifier, it is possible to obtain a highly sensitive piezoresistive effect in other semiconductors, as well as in metals.

\begin{figure*}[t!]
	\centering
	\includegraphics[width=5.4in]{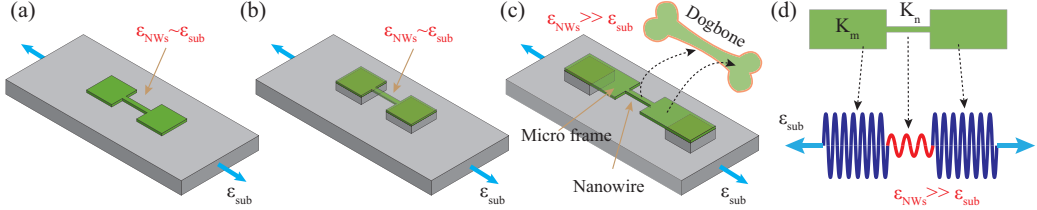}
		\vspace{-0.5em}
	\caption{Schematic sketches of nanowire strain sensors. (a)(b) Conventional non-released and released NW structure; 
		(c)(d) The proposed nano strain-amplifier and its simplified physical model.}
	\label{fig:fig1}
		\vspace{-1em}
\end{figure*}
Figure \ref{fig:fig1}(a) and 1(b) show the concept of the conventional structures of piezoresistive sensors. The piezoresistive elements are either released from, or kept on, the substrate. The sensitivity ($S$) of the sensors is defined based on the ratio of the relative resistance change ($\Delta R/R$) of the sensing element and the strain applied to the substrate ($\varepsilon_{sub}$):
\begin{equation}
S = (\Delta R/R)/\varepsilon_{sub}
\label{eq:sensitivity}
\end{equation}
In addition, the relative resistance change $\Delta R/R$ can be calculated from the gauge factor ($GF$) of the material used to make the piezoresistive elements: $\Delta R/R = GF \varepsilon_{ind}$, where $\varepsilon_{ind}$ is the strain induced into the piezoresistor. In most of the conventional strain gauges as shown in Fig. \ref{fig:fig1} (a,b), the thickness of the sensing layer is typically below a few hundred nanometers, which is much smaller than that of the substrate. Therefore, the strain induced into the piezoresistive elements is approximately the same as that of the substrate ($\varepsilon_{ind} \approx \varepsilon_{sub}$). Consequently, to improve the sensitivity of strain sensors (e.g. enlarging $\Delta R/R$), electrical approaches which can enlarge the gauge factor ($GF$) are required. Nevertheless, as aforementioned, the existence of the large gauge factor in nanowires due to quantum confinement or surface state, is still considered as controversial.  

It is also evident from Eq. \ref{eq:sensitivity} that the sensitivity of strain sensors can also be improved using a mechanical approach, which enlarges the strain induced into the piezoresistive element. Figure \ref{fig:fig1}(c) shows our proposed nano strain-amplifier structure, in which the piezoresistive nanowires are locally fabricated at the centre of a released bridge. The key idea of this structure is that, under a certain strain applied to the substrate, a large strain will be concentrated at the locally fabricated SiC nanowires. The working principle of the nano strain-amplifier is similar to that of the well-known dogbone structure, which is widely used to characterize the tensile strength of materials \cite{dogbone1,dogbone2}. That is, when a stress is applied to the dogbone-shape of a certain material, a crack, if generated, will occur at the middle part of the dogbone. The large strain concentrated at the narrow area located at the centre part with respect to the wider areas located at outer region, causes the crack. Qualitative and quantitative explanations of the nano strain-amplifier are presented as follows.  

For the sake of simplicity, the released micro frame and nanowire (single wire or array) of the nano strain-amplifier can be considered as solid springs, Fig. \ref{fig:fig1}(d). The stiffness of these springs are proportional to their width ($w$) and inversely proportional to their length (l): $K \propto w/l$. Consequently, the model of the released nanowire and micro frames can be simplified as a series of springs, where the springs with higher stiffness correspond to the micro frame, and the single spring with lower stiffness corresponds to the nanowire. It is well-known in classical physics that, for serially connected springs, a larger strain will be concentrated in the low--stiffness string, while a smaller strain will be induced in the high--stiffness string \cite{Springbook}. The following analysis quantitatively explained the amplification of the strain.	

\begin{figure}[b!]
	\centering
	\includegraphics[width=3in]{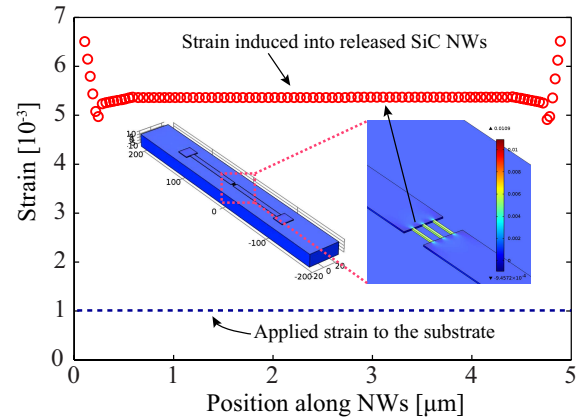}
	\vspace{-1em}
	\caption{Finite element analysis of the strain induced in to the nanowire array utilizing nano strain-amplifier.}
	\label{fig:fig2}
\end{figure}
When a tensile mechanical strain ($\varepsilon_{sub}$) is applied to the substrate, the released structure will also be elongated. Since the stiffness of the released frame is much smaller than that of the substrate, it is safe to assume that the released structure will follows the elongation of the substrate. The displacement of the released structure $\Delta L$ is:
\begin{equation}
\Delta L = \Delta L_m + \Delta L_n = L_m \varepsilon_m + L_n \varepsilon_n
\label{eq:displacement}
\end{equation} 
where $L_m$, $L_n$ are the length; $\Delta L_m$, $\Delta L_n$ are the displacement; and $\varepsilon_m$, $\varepsilon_n$ are the strains induced into the micro spring and nano spring, respectively. The subscripts m and n stand for the micro frames and nanowires, respectively. Furthermore, due to the equilibrium of the stressing force ($F$) along the series of springs, the following relationship is established: $F= K_m\Delta L_m = K_n \Delta L_n$, where $K_m$, $K_n$ are the stiffness of the released micro frames and nanowires, respectively. Consequently the relationship between the displacement of the micro frame (higher stiffness) and nanowires (lower stiffness) is:
\begin{equation}
\frac{\Delta L_m}{\Delta L_n}=\frac{K_n}{K_m}=\frac{L_mw_n}{L_nw_m}
\label{eq:euili}
\end{equation}
Substituting Eqn. \ref{eq:euili} into Eqn. \ref{eq:displacement}, the strain induced into the locally fabricated nanowires is:
\begin{equation}
\varepsilon_n = \frac{\Delta L_n}{L_n} = \frac{1}{1-\frac{w_m-w_n}{w_m}\frac{L_m}{L}}\varepsilon_{sub}
\label{eq:strainamp}
\end{equation} 

Equation \ref{eq:strainamp} indicates that increasing the ratio of $w_m/w_n$ and $L_m/L_n$ significantly amplifies the strain induced into the nanowire from the strain applied to the substrate. This model is also applicable to the case of nanowire arrays, in which $w_n$ is the total width of all nanowires in the array. 

The theoretical model is then verified using the finite element analysis (FEA). In the FEA simulation, we compare the strain induced into (i) non released nanowires, (ii) the conventionally released nanowires, and (iii) our nano strain-amplifier structure, using COMSOL Multiphysics \texttrademark. In our nano strain amplifying structure, the width of the released frame was set to be 8 $\mu$m, while the width of each nanowire in the array (3 wires) was set to be 370 nm. The nanowires array structure was selected as it can enhance the electrical conductance of the SiC nanowires resistor which makes the subsequent experimental demonstration easier. The ratio between the length of nanowires and micro bridge was set to be 1: 20. With this geometrical dimensions, strain induced into nanowires array $\varepsilon_n$ was numerically calculated to be approximately 6 times larger than $\varepsilon_{sub}$, Eqn. \ref{eq:strainamp}. The simulation results show that for all structure, the elongation of non-released and released nanowires follow that of the substrate. In addition, strain was almost completely transferred into conventional released and non-released structures. Furthermore, the ratio of the strain induced in to the locally fabricated nanowires was estimated to be 5.9 times larger than that of the substrate, Fig. \ref{fig:fig2}. These results are in solid agreement with the theoretical analysis presented above. For a nanowire array with an average width of 470 nm, the amplified gain of strain was found to be 4.5.   	

\begin{figure}[t!]
	\centering
	\includegraphics[width=3in]{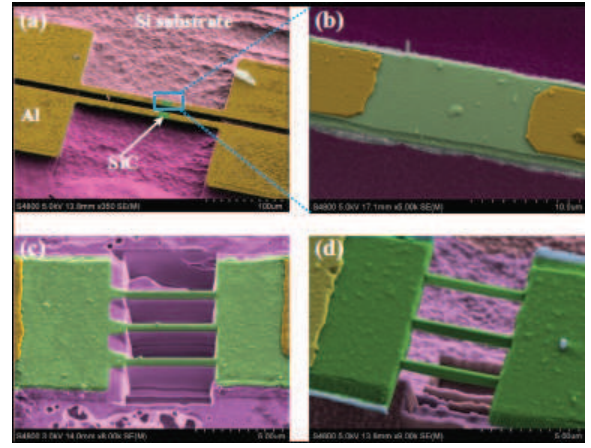}
	\caption{SEM image of SiC strain sensors. (a) Released SiC micro bridge used for the subsequent fabrication of the nano strain-amplifier; (b) SEM of a micro SiC resistor where the SiC nanowires array were formed using FIB; (c) SEM of non-released SiC nanowires; (d) SEM of locally fabricated SiC nanowires released from the Si substrate (nano strain-amplifier).}
	\label{fig:fig3}
	\vspace{-1em}
\end{figure}

Based on the theoretical analysis, we conducted the following experiments to demonstrate the high sensitivity of SiC nanowire strain sensors using the nano strain-amplifier. A thin 3C-SiC film with its thickness of 300 nm was epitaxially grown on a 150 mm diameter Si wafer using low pressure chemical vapour deposition \cite{SiC_growth}. The film was \emph{in situ} doped using Al dopants. The carrier concentration of the p-type 3C-SiC was found to be $5 \times 10^{18}$ cm$^{-3}$, using a hot probe technique \cite{philip}. The details of the characteristics of the grown film can be found elsewhere \cite{Phan_JMC}. Subsequently, I-shape p-type SiC resistors with aluminum electrodes deposited on the surface were patterned using inductive coupled plasma (ICP) etching. As the piezoresistance of p-type 3C-SiC depends on crystallographic orientation, all SiC resistors of the present work were aligned along [110] direction to maximize the piezoresistive effect. Next, the micro scale SiC resistors were then released from the Si substrate using dry etching (XeF$_2$). Finally, SiC nanowire arrays were formed at the centre of the released bridge using focused ion beam (FIB). Two types of nanowire array were fabricated with three nanowires for each array. The average width of each nanowire in each type were 380 nm and 470 nm, respectively. Figure \ref{fig:fig3} shows the SEM images of the fabricated samples, including the conventional released structure, non-released nanowires, and the nano strain-amplifier.  

\begin{figure}[b!]
	\centering
	\includegraphics[width=3in]{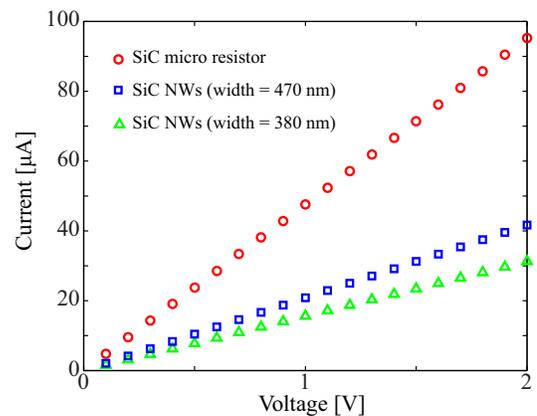}
		\vspace{-1.5em}
	\caption{Current voltage curves of the fabricated SiC resistors.}
	\label{fig:IV}
\end{figure}

The current voltage (I-V) curves of all fabricated samples were characterized using a HP 4145 \texttrademark ~parameter analyzer. The linear relationship between the applied voltage and measured current, indicated that Al made a good Ohmic contact with the highly doped SiC resistance, Fig. \ref{fig:IV}. Additionally, the electrical conductivity of both nanowires and micro frame estimated from the I-V curve and the dimensions of the resistors shows almost the same value. This indicated that the FIB process did not cause a significant surface damage to the fabricated nanowires.    
	
The bending experiment was used to characterize the piezoresistive effect in micro size SiC resistors and locally fabricated SiC nanowire array. In this experiment one end of the Si cantilever (with a thickness of 625 $\mu$m, and a width of 7 mm) was fixed while the other end was deflected by applying different forces. The distance from the fabricated nanowires to the free end of the Si cantilever was approximately 45 mm. The strain induced into the Si substrate is $\varepsilon_\text{sub} = Mt/2EI$, where $M$ is the applied bending moment; and $t$, $E$ and $I$ are the thickness, Young's modulus and the moment of inertia of the Si cantilever, respectively. The response of the SiC resistance to applied strain was then measured using a multimeter (Agilent \texttrademark 34401 A). 

\begin{figure}[h!]
	\centering
	\includegraphics[width=3in]{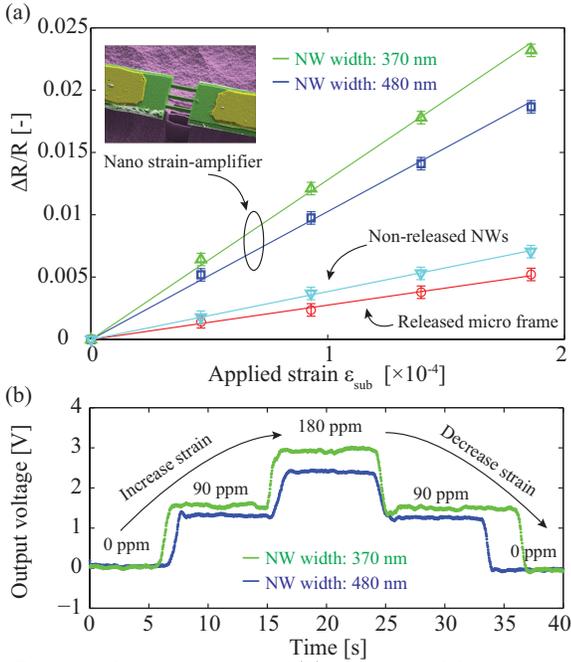}
		\vspace{-1.5em}
	\caption{Experimental results. (a) A comparision between the relative resistance change in the nano strain-amplifiers, non released nanowires and released micro frames; (b) The repeatability of the SiC nanowires strain sensors utilizing the proposed structure.}
	\label{fig:DRR}
			\vspace{-1em}
\end{figure}	
The relative resistance change ($\Delta R/R$) of the micro and nano SiC resistors was plotted against the strain induced into the Si substrate $\varepsilon_{sub}$, Fig. \ref{fig:DRR}(a). For all fabricated samples, the relative resistance change shows a good linear relationship with the applied strain ($\varepsilon_{sub}$). In addition, with the same applied strain to the Si substrate, the resistance change of the SiC nanowires using the nano strain-amplifier was much larger than that of the the SiC micro resistor and the conventional non-released SiC nanowires. In addition, reducing the width of the SiC nanowires also resulted in the increase of the sensitivity. The magnitude of the piezoresistive effect in the nano strain-amplifier as well as conventional structures were then quantitatively evaluated based on the effective gauge factor ($GF_{eff}$), which is defined as the ratio of the relative resistance change to the applied strain to the substrate: $GF_{eff} = (\Delta R/R)/\varepsilon_{sub}$. Accordingly, the effective gauge factor of the released micro SiC was found to be 28, while that of the non-released SiC nanowires was 35. From the data shown in Fig. \ref{fig:DRR}, the effective gauge factor of the 380 nm and 470 nm SiC nanowires in the nano strain-amplifier were calculated as 150 and 124, respectively. Thus for nanowire arrays with average widths of 380 nm and 470 nm, the sensitivity of the nano strain-amplifier was 5.4 times and 4.6 times larger than the bulk SiC, respectively. These results were consistent with analytical and numerical models presented above. The relative resistance change of the nano strain-amplifier also showed excellent linearity with the applied strain, with a linear regression of above 99\%. 

The resistance change of the nano strain-amplifier can also be converted into voltage signals using a Wheatstone bridge, Fig. \ref{fig:DRR}(b). The output voltage of the nano strain-amplifier increases with increasing tensile strains from 0 ppm to 180 ppm, and returned to the initial value when the strain was completely removed, confirming a good repeatability after several strain induced cycles. The linearity of the relative resistance change, and the repeatability indicate that the proposed structure is promising for strain sensing applications.
    
In conclusion, this work presents a novel mechanical approach to obtain highly sensitive piezoresistance in nanowires based on a nano strain-amplifier. The key factor of the nano strain-amplifier lies on nanowires locally fabricated on a released micro structure. Experimental studies were conducted on SiC nanowires, confirming that by utilizing our nano strain-amplifier, the sensitivity of SiC nanowires was 5.4 times larger than that of conventional structures. This result indicated that the nano strain-amplifier is an excellent platform for ultra sensitive strain sensing applications.    


\end{document}